\documentclass[fleqn,10pt]{wlscirep}
\usepackage[utf8]{inputenc}
\usepackage[T1]{fontenc}
\usepackage{lineno}
\usepackage{siunitx}
\usepackage{hyperref}
\nolinenumbers

\title{2DeteCT - A large 2D expandable, trainable, experimental Computed Tomography dataset for machine learning}

\author[1,*]{Maximilian B. Kiss}
\author[1,2]{Sophia B. Coban}
\author[1,3]{K. Joost Batenburg}
\author[1,4]{Tristan van Leeuwen}
\author[1,*]{Felix Lucka}
\affil[1]{Centrum Wiskunde \& Informatica, Computational Imaging group, Amsterdam, 1098 XG, The Netherlands}
\affil[2]{Department of Mathematics, University of Manchester, Oxford Road, Manchester, M13 9PL. United Kingdom}
\affil[3]{Leiden University, LIACS, Leiden, 2300 RA, The Netherlands}
\affil[4]{Utrecht University, Mathematical Institute, Utrecht, 3584 CD, The Netherlands}

\affil[*]{corresponding author(s): Maximilian B. Kiss (maximilian.kiss@cwi.nl), Felix Lucka (felix.lucka@cwi.nl)}

\begin{abstract}
Recent research in computational imaging largely focuses on developing machine learning (ML) techniques for image reconstruction, which requires large-scale training datasets consisting of measurement data and ground-truth images. However, suitable experimental datasets for X-ray Computed Tomography (CT) are scarce, and methods are often developed and evaluated only on simulated data. We fill this gap by providing the community with a versatile, open 2D fan-beam CT dataset suitable for developing ML techniques for a range of image reconstruction tasks. To acquire it, we designed a sophisticated, semi-automatic scan procedure that utilizes a highly-flexible laboratory X-ray CT setup. A diverse mix of samples with high natural variability in shape and density was scanned slice-by-slice (5000 slices in total) with high angular and spatial resolution and three different beam characteristics: A high-fidelity, a low-dose and a beam-hardening-inflicted mode. In addition, 750 out-of-distribution slices were scanned with sample and beam variations to accommodate robustness and segmentation tasks. We provide raw projection data, reference reconstructions and segmentations based on an open-source data processing pipeline.\\
\end{abstract}
\begin{document}

\flushbottom
\maketitle

\thispagestyle{empty}

\section*{Background \& Summary}
\label{sec:background}
X-ray computed tomography (CT) is a non-invasive X-ray absorption-based imaging technique used in a range of fields, including medicine, manufacturing industry, food industry, and materials science. For a CT scan, X-ray projection images of an object are taken from multiple angular positions. To obtain a reconstruction of this acquired data, an inverse problem has to be solved through analytical methods such as filtered back-projection or iterative reconstruction algorithms. The overarching methodology to reconstruct images from these measurements is called computational imaging.
\\[6pt]
In recent years, the field of computational imaging focused on developing data-driven methods for image reconstruction \cite{Ravishankar_2019}. With that focus also the need for large datasets increased. In particular, image reconstruction based on deep learning (DL) methods, such as deep neural networks (DNNs), requires a large amount of realistic data for both evaluating the developed methods on real world applications as well as for constructing the method itself. For example, supervised learning approaches optimize the network parameters based on training data composed of a large number of representative pairs of input and desired ideal output data of the network (i.e., the ground truth).
\\[6pt]
While in various application fields of DL there already exist large and open data collections such as MNIST \cite{MNIST_1998} for handwritten digit recognition or IMAGENET for image classification/processing \cite{IMAGENET_2009}, suitable experimental data collections for computational imaging with high-quality ground truth reconstructions and/or segmentations are scarce. For magnetic resonance imaging (MRI), the fastMRI dataset \cite{Knoll_2020} is a larger dataset containing raw (unprocessed) k-space data of knees and brains acquired across multiple institutions and scanners. There are some datasets available for X-ray CT but unfortunately they lack certain desirable characteristics: The Mayo clinic low-dose CT challenge of 2016 \cite{mccollough2016tu} with 30 patient scans consisting of roughly 70 slices each has a fairly small number of scan subjects. Although their new release of 2021 \cite{Mayo_2021} has 300 patients another important downside of both datasets is that noisy reconstruction and projection data is simulated from clean reconstructed volumes. The LoDoPaB-CT dataset \cite{leuschner2021lodopab} contains over 40,000 scan slices from around 800 patients selected from the LIDC/IDRI database. But despite the large size of the data collection it still uses simulated low photon count measurements and not experimental data. The walnut dataset \cite{Sarkissian_2019} provides 42 three-dimensional cone-beam CT (CBCT) scans of walnuts. Although it provides raw experimental data, the applicability of the dataset is limited through the small number of samples of the same object type and its design for a specific task in 3D CBCT. This makes it less useful for more general methods development. Overall, the few available datasets are limited in their applicability to one computational imaging task.
\\[6pt]
A key disadvantage of available datasets for X-ray CT is that they commonly use commercial CT solutions with licensed software that have no (or limited) access to raw projection data or the specifics of the experimental acquisition. Therefore, mathematical and computational studies typically rely on artificial data simulated with varying degrees of realism. To develop and train algorithms for computational imaging tasks such as low-dose reconstruction, limited or sparse angular sampling, beam-hardening artifact reduction, super-resolution, region-of-interest tomography or segmentation it is necessary to have corresponding experimental training data. The field of X-ray CT still lacks such a large-scale, versatile, experimental dataset for machine learning. Especially two-dimensional, reconstructed CT slices would be useful for method development since the corresponding learning and reconstruction tasks require less computational resources compared to their three-dimensional counterparts.
\\[6pt]
Acquiring such a large 2D CT dataset encompasses various requirements: First, research groups need to have a scanning facility readily available and be able to also use it for a large-scale, time-extensive data collection process. Second, the geometry and other acquisition parameters of this scanner must be highly adjustable to collect a dataset that can be used for a wide range of machine learning applications. Third, similar image characteristics as encountered in medical CT would be preferable because of the great importance of medical imaging as an application area of X-ray CT means. Last, it is necessary to limit manual intervention during the acquisition process to be able to reach a high number of acquired CT reconstruction slices. This requires the ability to automatize the acquisition process as much as possible.
\\[6pt]
In this paper, we describe in detail the steps involved in acquiring an unprecedented X-ray data collection by making extensive use of a highly flexible, programmable and custom-built X-ray CT scanner: The first step was to choose suitable scan parameters for the acquisition such as beam filtration, X-ray tube voltage and current, detector exposure time, binning and averaging, the number of projection angles and source, object and detector positions. The aim behind these choices was to acquire a rich projection dataset for each image slice that can be used for a wide range of imaging tasks such as supervised or unsupervised denoising, sparse-angle scanning, beam-hardening reduction, super-resolution, region-of-interest tomography or segmentation. As a second step, a scanning object had to be designed in such a way that 2D slice scans resemble image features found in medical abdominal CT scans. For this a cylindrical tube was filled with a mix of samples of similar density but different shapes immersed in a powder. In the third step, an experimental set-up and a script generator program was developed that allowed to automatize the collection of 50 slices during an 8.5h scan. In 111 scanning sessions (each with a different sample mix) and a total scanning time of more than 850 hours we acquired 5,000 slices over a duration of almost five months. Each of these slices was acquired in three different acquisition modes resulting in "clean", "noisy", and "artifact-afflicted" reconstructions. Furthermore, additional 750 out-of-distribution (OOD) slices were acquired, for which each ingredient was scanned separately, and scans with different parameters and/or samples were designed. This was done to accommodate robustness tests or help with segmentation / foreign object detection tests.

\begin{figure}[!ht]
	\centering
	\includegraphics[width=0.75\columnwidth]{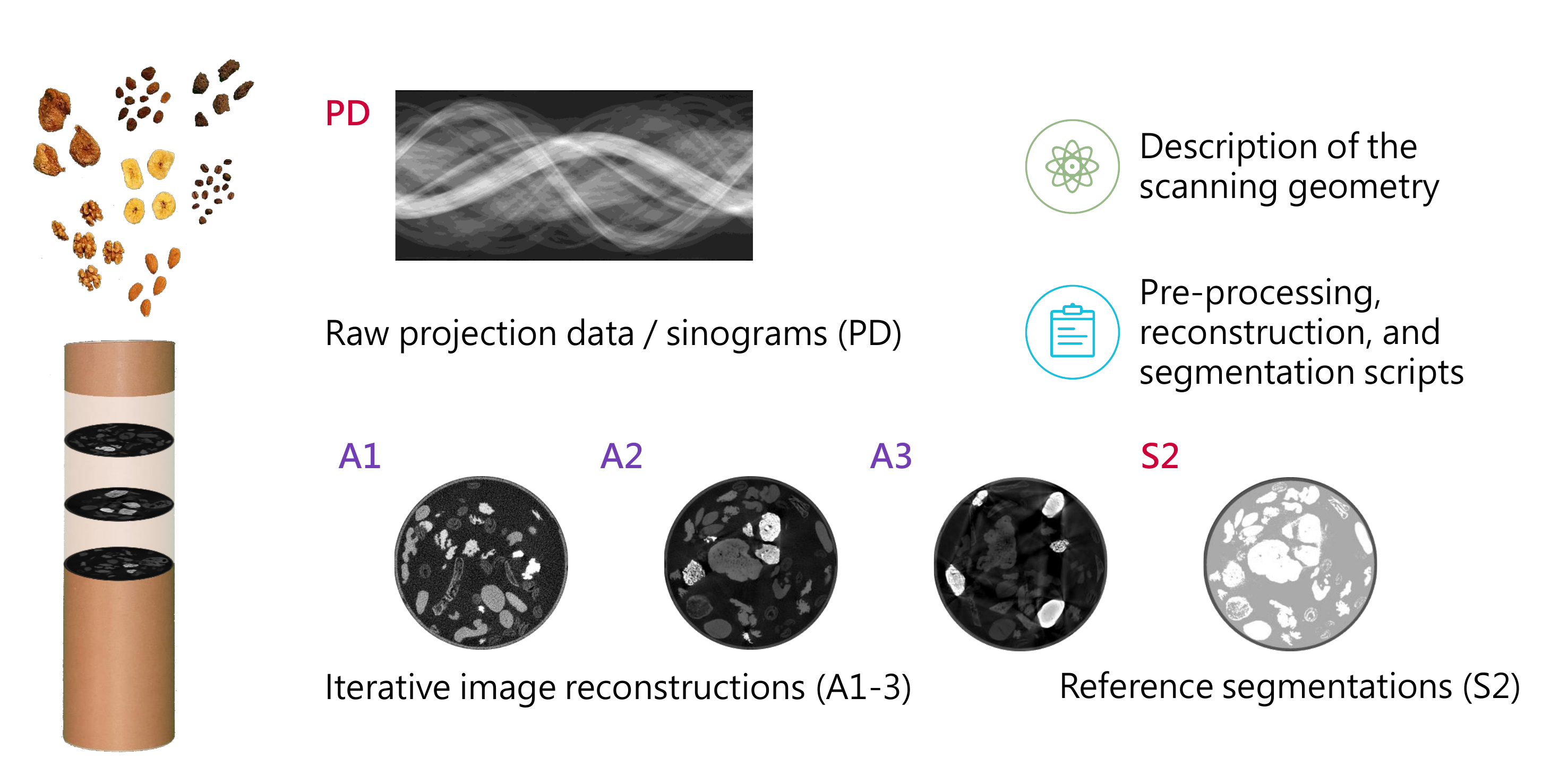}
	\caption{Overview of the scope of the 2DeteCT dataset.}
	\label{fig:teaser_figure}
\end{figure}

To make this dataset accessible to a broad range of researchers including those who do not have high-performance computing facilities readily available, we provide also reference reconstructions and segmentations, additionally to the raw projection data in sinograms and an implementation of the complete computational pipeline based on open-source software (cf. Figure \ref{fig:teaser_figure}).
\\[6pt]
The structure of the paper is as follows: The section "Experimental design" describes the overall study design, the experimental design of the acquisition and data processing protocols. In the section "Data records" the file structure of the data collection is described, while the "Usage notes" section contains all details on how to access and use it.

\section*{Methods}
\label{sec:methods}
\subsection*{X-ray computed tomography scanner} \label{subsec:FleXray}
The data collection has been acquired using a highly flexible, programmable and custom-built X-ray CT scanner, the FleX-ray scanner\cite{FleXrayLab}, developed by TESCAN-XRE NV (\href{https://info.tescan.com/micro-ct}{https://info.tescan.com/micro-ct}), located in the FleX-ray Lab at the Centrum Wiskunde \& Informatica (CWI) in Amsterdam, Netherlands. It consists of a cone-beam microfocus X-ray point source (limited to 90 kV and 90 W) that projects polychromatic X-rays onto a 14-bit CMOS (complementary metal-oxide semiconductor) flat panel detector with CsI(Tl) scintillator (Dexella 1512NDT, \cite{Detector}) and $1536 \times 1944$ pixels, $\SI{74.8}{\micro\metre^2}$ each. To create a 2D dataset, a fan-beam geometry was mimicked by only reading out the central row of the detector. Between source and detector there is a rotation stage, upon which samples can be mounted, cf. Figure \ref{fig:Inside-Scanner}. The machine components (i.e., the source, the detector panel, and the rotation stage) are mounted on translation belts that allow the moving of the components independently from one another. Furthermore, we developed an in-house software toolbox for designing executable scan scripts for the scanner. With these, the scanner performs sophisticated acquisition protocols automatically without human intervention. The acquisition procedure that we programmed with this toolbox will be described in more detail in the "Data acquisition" section.

\subsection*{Experimental design}
\label{sec:experimental_design}
The three general aims of this data collection are as follows. First, to provide the computational imaging community with a dataset that encompasses raw experimental measurement data that can be used to develop and test techniques in CT imaging for real world applications. These also include medical CT by producing similar image features and contrast in the dataset slices as exhibited in medical abdominal CT scans. Second, the possibility to expand the current scope of the dataset by adding more detailed multi-class segmentations or by adding more slices with the same or a different sample mix because of a reproducible setup. Third, to have a large enough dataset such that it can be used for developing deep learning algorithms for different computational imaging applications, including low-dose acquisition, limited or sparse-angle scanning, beam-hardening artifact reduction, super-resolution, region-of-interest tomography or segmentation. The three key features to achieve this were firstly, the design of a semi-automatic data acquisition; secondly, finding a scanning object and sample mix that is both, diverse enough as well as stable over the long scanning time; and thirdly, creating a scanning setup and experimental design that enables the aforementioned applications. A total of 9 months went into the experimental design, developing and scripting the semi-automatic data acquisition, selecting and testing the different samples, designing the scanning setup and determining suitable acquisition parameters.

\subsubsection*{Semi-automatic data acquisition}
\label{sec:semi_automatic}
To maximize the number of scanned CT slices it was strictly necessary to limit the amount of human intervention during the data acquisition. The idea was to use the flexibility of the scanner and its ability to be scripted to automatize the acquisition process as much as possible. The only necessary human interaction would be to prepare the next sample mix and to start a scanning protocol, which then would acquire a certain number of slices automatically. In our optimized setting, acquiring a batch of 10 slices takes a fixed time of 1h 42mins, while 50 slices can be acquired in 8h 34mins. This way, the idle time of the scanner, e.g. at night, could be used to reach a large amount of slices.

\subsubsection*{Sample preparation}
\label{sec:sample}
One of the aims of this dataset was to produce images with similar image features and contrast as abdominal medical CT scans. To achieve this, a container representing the body was used as a scanning object and was filled with a mix of cm-scale objects mimicking the organs/bones submersed in a background medium resembling connective tissue. Furthermore, this mix of sample objects should have a high natural variability in both inter and intra-sample shape and density (see Table \ref{tab:densities_sample_mix}). In particular, one of the samples should be dense enough to correspond to bones/teeth and introduce beam-hardening effects. Lastly, this sample mix should stay stable during long lasting high-intensity X-ray exposure and therefore have especially a certain temperature stability.
\\[6pt]
To fulfill these requirements a variety of dried fruits and nuts were scanned and their appearance as well as their relative intensities in the reconstructed image slices were evaluated. Although the actual attenuation coefficients of the samples are not the same as for organs, bones and connective tissue in medical CT scans, the ratio between dense and soft regions are similar. Furthermore, some of the dried fruits and nuts have shapes that resemble organs, e.g. walnuts resembling brain tissue. The tested samples included: dried apricots, bananas, dates, figs, mangoes, sultanas, and coffee beans as well as almonds, cashews, hazelnuts, para nuts, peanuts, pecans, pistachios, and walnuts. Furthermore, different stone types were tested for the sample selection as objects that introduce beam hardening effects such as different types of basalt, granite, lime stone, lava stone, marble, quartz, and slate. To avoid air volumes between the samples, various filler materials were investigated: cereal-based coffee powder, sand, saw dust, (powdered) sugar, salt, and sweetener. The requirements were similar or lower density than water, contrast to other samples, no/limited air bubbles, not too coarse in its fine structure, rather homogeneous, and temperature stable regarding aggregate state, physical extension, as well as density/humidity.
\\[6pt]
After evaluating all sample mix tests, the following sample mix was selected: almonds, dried banana chips, coffee beans, dried figs, lava stones, raisins, and walnuts immersed in cereal-based coffee powder as a filler material. The respective densities of these samples can be found in the appendix in Table \ref{tab:densities_sample_mix}. Since X-ray absorption is related to the density and thickness of the scanned material, these values are a first indicator for the measured intensities and their contrast to each other.
\\[6pt]
To avoid the samples to dry up too much over time through the long exposure to high energy X-ray radiation, the sample mix was replaced three times in total. The amount of each sample within the final mixes can be found in Table \ref{tab:sample_mixes} and was based on yielding a roughly equal share of volume within the sample mix.
\\[6pt]
The aforementioned sample mix was put into a sample container as a scanning object. The requirements for this were apart from a stable positioning and temperature/radiation exposure stability that the container did not absorb too much of the X-ray radiation. After testing different paper and plastic (PE, PVC, etc.) based containers a cylindrical cardboard tube was selected.
\\[6pt]
Because of the restrictions imposed by the inner dimensions of the scanner and the maximal diameter usable on the scanner sample stage (109.4mm) a cardboard tube with 10cm inside diameter and 34cm of height was used. To prevent unwanted scattering from the aluminum sample stage the sample tube was elevated by a few centimeters with a 3D-printed PETG sample stage cylinder of 50.0mm in height. To ensure that the setup remains the same this cylinder can be steadily and reproducibly positioned into a carved out circle on the sample stage and the cardboard tube was positioned centrally onto this sample stage cylinder using superglue. 
\\[6pt] 
For the additional 750 out-of-distribution (OOD) slices, each ingredient in the mix was scanned separately and new samples were included to accommodate robustness tests or help with segmentation or foreign object detection tests. The OOD sample objects were fresh figs, grapes, hazelnuts, pistachios, peanuts and titanium prostheses screws. The latter were chosen to have objects in the sample mix that create even more severe artefacts than the lava stones and furthermore are used in clinical practice.

\begin{figure}[!ht]
	\centering
	\includegraphics[width=0.75\columnwidth]{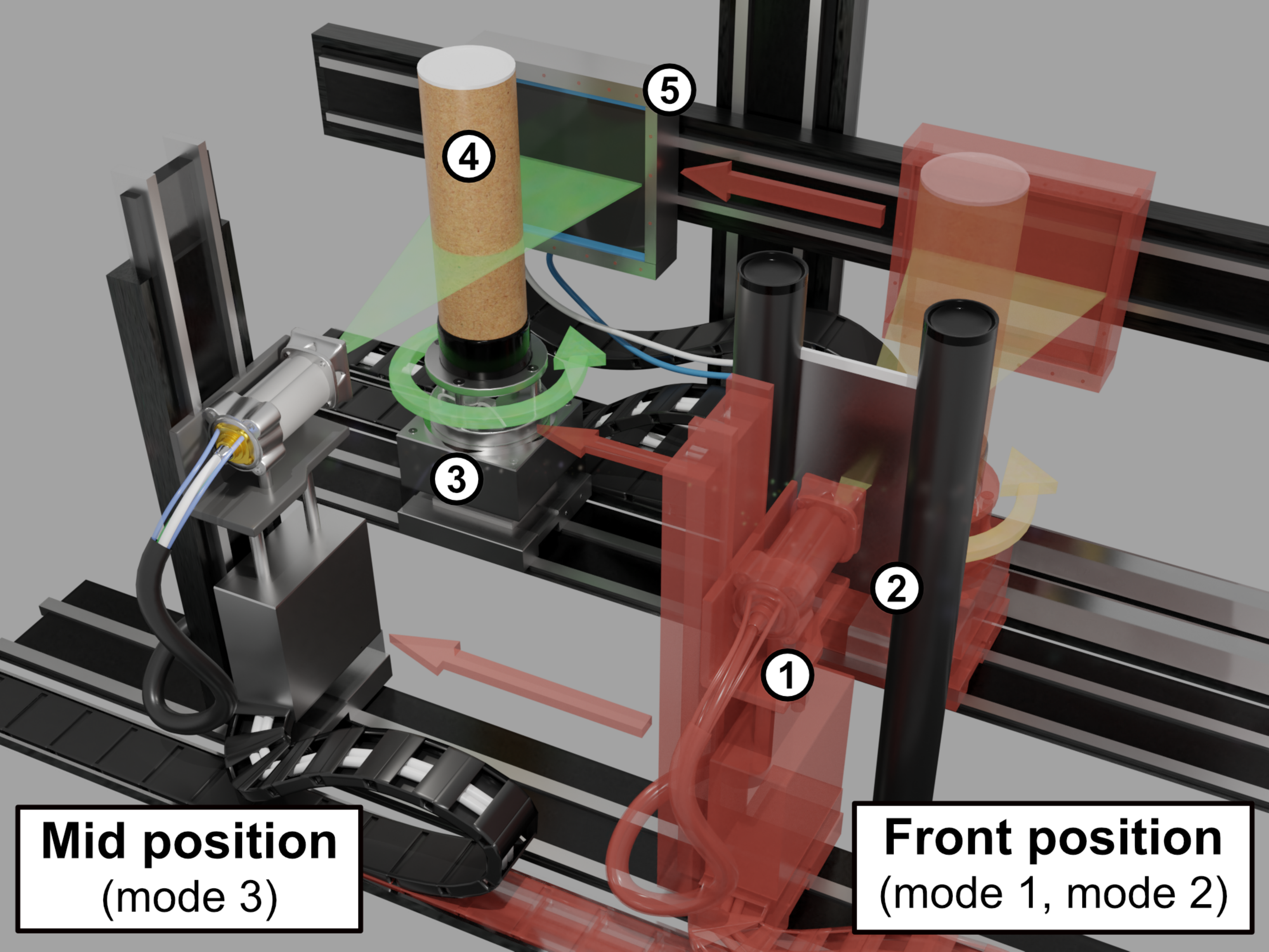}
	\captionsetup{format=hang}
	\caption{FleX-ray Lab: the computed tomography set-up used for the data acquisition. 1) cone-beam X-ray source; 2) Thoraeus filter sail; 3) Rotation stage; 4) Sample tube; 5) Flat panel detector. The objects 1, 3, 4, and 5 move from their red transparent front position to the mid position for the acquisitions of mode 3. In both positions 3,601 projection images per slice are taken while the object rotates 360 degrees.}
	\label{fig:Inside-Scanner}
\end{figure}

\subsubsection*{Scanning setup and parameter choices}
\label{sec:scanning_setup}
The scanning setup had to be designed in such a way that the different application areas for the data collection could be served. Namely, denoising, sparse-angle scanning, beam-hardening reduction, super-resolution, region-of-interest tomography or segmentation. Since data for sparse-angle, super-resolution and region-of-interest tomography can be generated from scans with a large amount of angle projections and high resolution, the objectives were to acquire at least a noisy, a beam-hardening artifact-inflicted and a "clean" scan as a ground truth and starting point for constructing a high-confidence segmentation.
\\[6pt]
Therefore, data has to be acquired in three different acquisition modes. To achieve both beam-hardening artifact-inflicted and "clean" images from the same sample it is necessary to scan with and without beam filters. Since the FleX-ray scanner does not encompass an automatic filter wheel we developed a new kind of setup within our scanner cabinet. To be able to scan multiple vertical slices with a filter we built a "filter sail" which was placed between the X-ray tube and the sample tube. Since the source, rotation stage and detector can be moved at the same time it was possible to get this filter sail into the beam axis without human interaction during the process (cf. red arrows in Figure \ref{fig:Inside-Scanner}).
\\[6pt]
For this a second scanning position in the scanning cabinet was identified which is located in the front instead of the middle of the scanner (on the transversal axis). After testing that this scanning position yields indistinguishable results with otherwise identical acquisition parameters to the "mid position" a scanning script was developed that will move between these two position depending on which acquisition mode shall be used at that moment.
\\[6pt]
The "front position" was used for mode 1 and 2 whereas mode 3 was acquired in the "mid position". This distinction was necessary since mode 1 and 2 require a filter setup while measurements in mode 3 are acquired without a filter. This means in the "front position" the X-rays go through the filter sail whereas in the "mid position" they are not filtered. To limit the time spent on motor movements the scans were carried out in batches of 10 slices in each mode.
\\[6pt]
The next choice for the scanning setup regarded the positioning within the scanner cabinet and encompassed three objectives: Sufficient photon flux, high resolution, and good detector coverage. These are mainly influenced by two parameters, the so-called Source-to-Object Distance (SOD) and Source-to-Detector Distance (SDD). The bigger the SDD is, the more parallel the beam geometry is. At the same time though the photon flux decreases. While it is desirable to have a parallel beam geometry, a decrease in photon flux necessitates longer scanning times or the noise increases which prolongs the data acquisition significantly. The SOD and its ratio with the SDD, called magnification factor $mag=\frac{SDD}{SOD}$, determine the resolution of the scan. This means what length inside the object is covered by one detector pixel $det$ and accordingly, the resolution can be calculated as follows: $res = \frac{det}{mag}$. We strived to maximize spatial resolution subject to the constraint that the size of the scanned object does not exceed the size of the detector.
\\[6pt]
The positioning of both sample tube and detector was limited by the dimensions of the inside of the scanner, the sample tube, the size of the detector, and the possible motor movements. With a minimal motor position distance between detector and rotation stage of 63mm on the magnification axis, 529mm was found to be a suitable SDD and increasing the distance to the sample tube to an SOD of 431mm increased the footprint of the scanned object with respect to the width of the detector. This resulted in a sufficient photon flux and a resolution of $\SI{60.95}{\micro\metre^3}$.
\\[6pt]
After the selection of the sample mix and sample tube as well as the positioning suitable beam parameters for the desired application areas had to be found. Two main problems arise though for the chosen scanning setup and experimental design: Firstly, lab X-ray sources emit a spectrum of X-ray energies. They have a so-called polychromatic beam in contrast to synchroton facilities which have a monochromatic beam consisting only of X-rays of one distinct energy \cite{als2011elements}. The broader the beam spectrum is, the larger the occurring beam hardening effects are \cite{Patton_Turkington, VandeCasteele_2002}. Secondly, low energy photons are absorbed more strongly by larger objects limiting the amount of detected photons. Therefore, a low average beam energy leads to high noise. Hence, the beam spectrum of the X-rays needed to be optimized to produce "clean" reconstructions with limited noise and beam-hardening artifacts. 
\\[6pt]
For this different combinations of tube voltage and filters to shape the beam spectrum were tested. First, the lower bound of the X-ray energy required to penetrate the object was determined. A tube voltage of 40.0kV produced virtually no signal on the detector and the noise was too high. For a tube voltage of 60.0kV and maximum current of $\SI{1000}{\micro\ampere}$ a sufficient amount of photons was measured, but without using filters noticeable beam hardening artifacts were observed in the reconstructions. Next, the tube voltage was set to the maximum of 90.0kV and beam filtration was used to improve the image quality. To reduce the beam hardening effects the low-energy photons were filtered out by placing thin sheets of metal between the X-ray tube and the sample tube. A variety of filters of different materials (Al, Cu, Sn, W) and thicknesses (0.01mm - 0.50mm) as well as combinations of them were tested. A popular compound filter in CT imaging, the so-called "Thoraeus filter" \cite{khan2014khan}, showed the best performance. Its compound consists of a tin filter, followed by a copper and after that a aluminum filter of varying thicknesses and effectively reduces the amount of photons carrying an energy of 1.5keV to 70.0keV. After testing a variety of thicknesses for the different compounds the final filter setup was composed of Sn = 0.1mm, Cu = 0.2mm, and Al = 0.5mm. To have sufficient signal the tube current had to be set again to the maximum of $\SI{1000}{\micro\ampere}$. Figure \ref{fig:beam_spectrum} illustrates that with the chosen compound filter almost all photons with energies below 40.0keV are filtered out of the beam spectrum which will reduce the beam hardening artifacts:

\begin{figure}[!btp]
	\centering
	\includegraphics[width=0.75\columnwidth]{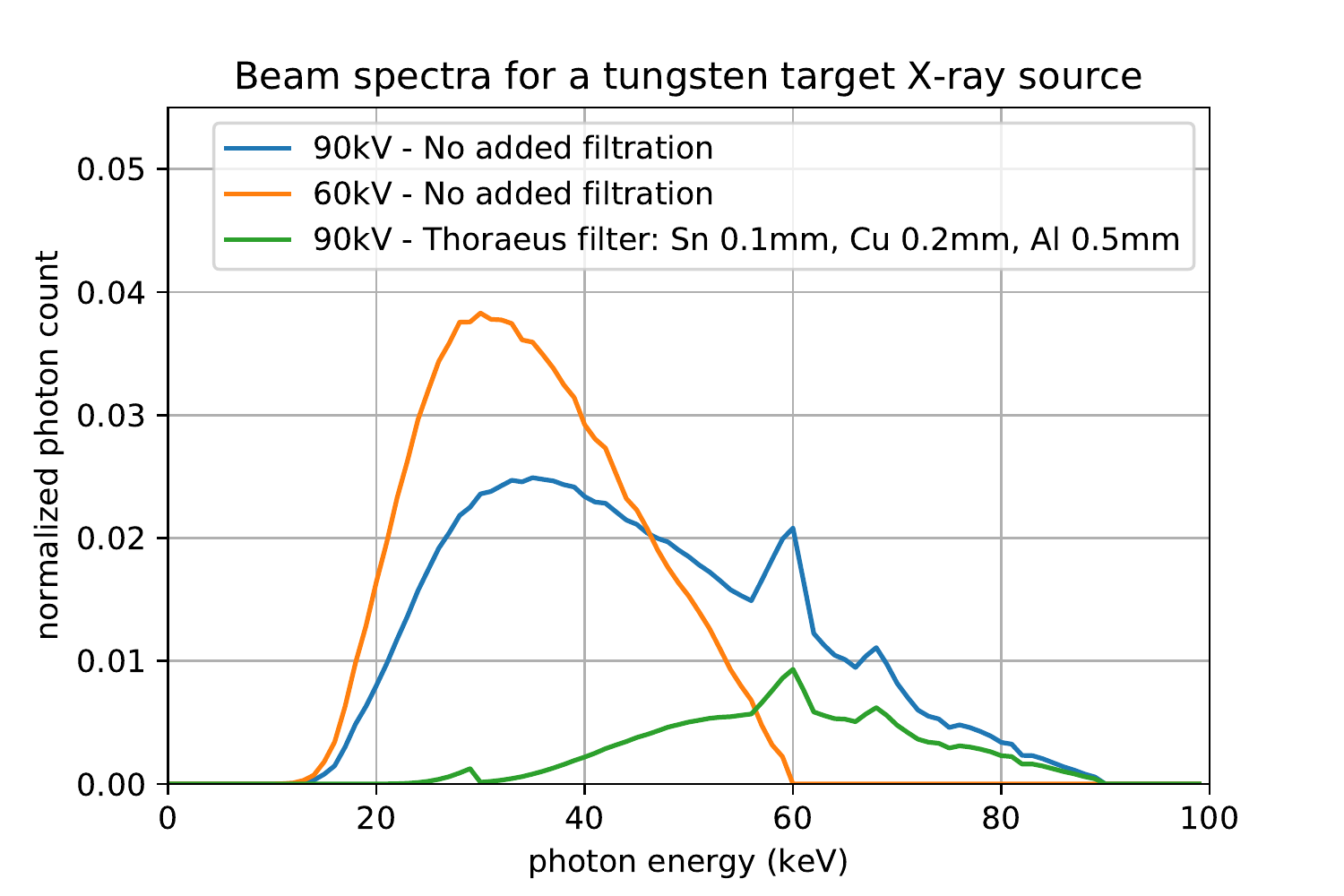}
	\caption{Beam spectra of a tungsten target X-ray source with an X-ray exit window made of \SI{300}{\micro\meter} Beryllium operated at 60kV with no added filtration and 90kV tube voltage with no added filtration and filtered with a Thoraeus filter of Sn = 0.1mm, Cu = 0.2mm, Al = 0.5mm simulated by TASMIP software \cite{TASMIP}.}
	\label{fig:beam_spectrum}
\end{figure}

Lastly, the exposure time and the number of projections for the acquisition of the scans was chosen. For the latter an application of the Nyquist-Shannon sampling theorem to CT yields that the achievable image resolution is not limited by the angular sampling rate if the number of projections is chosen greater than the number of detector pixels times $\pi/2$\cite{Kharfi_2013}, which amounts to $1912\times \frac{\pi}{2}=3003$ in our setting. Therefore, 3601 projections were chosen where the first and last projection coincide to have a standard angular increment of $0.1\deg$. For the exposure time 50ms (20Hz) ensured that all projections acquired for one slice are obtained within 3min scanning time without saturating the detector in any of the acquisition modes. All projection images were taken without any hardware binning or averaging.

\begin{table}[ht]
\centering
\begin{tabular}{|c|c|c|c|c|c|}
\hline
\textbf{Acquisition parameter} & \textbf{Mode 1} & \textbf{Mode 2} & \textbf{Mode 3} & \textbf{Noise-OOD} & \textbf{BH-OOD *$^1$}\\
\hline
Tube voltage & $\SI{90.0}{\kilo\volt}$ & $\SI{90.0}{\kilo\volt}$ & $\SI{60.0}{\kilo\volt}$ & $\SI{90.0}{\kilo\volt}$ & $\SI{45.0}{\kilo\volt}$ \\
\hline
Tube power & $\SI{3.0}{\watt}$ & $\SI{90.0}{\watt}$ & $\SI{60.0}{\watt}$ & $\SI{1.5}{\watt}$ & $\SI{45.0}{\watt}$\\
\hline
Filters used & Thoraeus*$^2$ & Thoraeus & No Filter & Thoraeus & No Filter \\
\hline
Exposure time & \multicolumn{4}{c|}{$\SI{50.0}{\milli\second}$} & $\SI{50.0}{\milli\second}$ / $\SI{110.0}{\milli\second}$ \\
\hline
Effective detector pixel size & \multicolumn{5}{c|}{$\SI{74.8}{\micro\metre}$} \\
\hline
Source to object distance *$^3$ & \multicolumn{5}{c|}{$\SI{431.020}{\milli\metre}$} \\
\hline
Source to detector distance *$^3$ & \multicolumn{5}{c|}{$\SI{529.000}{\milli\metre}$} \\
\hline
Number of projections & \multicolumn{5}{c|}{3601} \\
\hline
Angular increment & \multicolumn{5}{c|}{$\SI{0.1}{\deg}$} \\
\hline
\end{tabular}
\caption{\label{tab:acquisition_parameters}Summary of the acquisition parameters used. *$^1$(BH = beam hardening), *$^2$(Thoraeus = Sn 0.1mm, Cu 0.2mm, Al 0.5mm), *$^3$ these quantities are based on the motor readings of the FleX-ray scanner which get translated into physical quantities and are subject to alignment errors.}
\end{table}

\subsection*{Data acquisition}
\label{sec:data_acquisition}
\begin{figure}[!ht]
	\centering
	\includegraphics[width=0.90\columnwidth]{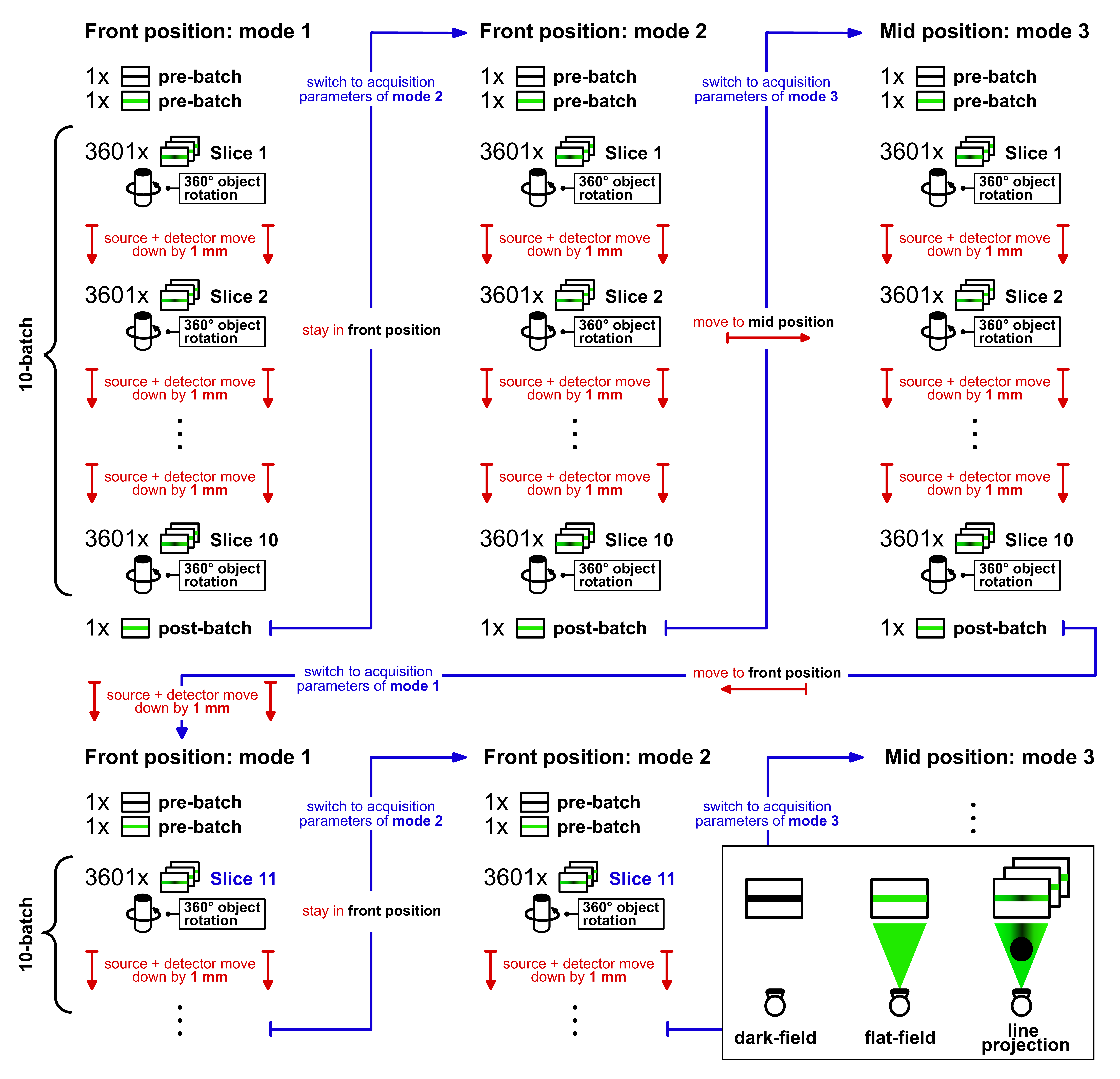}
	\caption{Visualization of the scanning procedure}
	\label{fig:scanning_procedure}
\end{figure}

As described in the section "Experimental design" the data acquisition was done in a semi-automatic fashion. Using our in-house script generator (cf. Section "X-ray computed tomography scanner") we developed a scan protocol that can acquire 50 slices in all three acquisition modes in one continuous session lasting 8h 34mins. The most time consuming processes in this acquisition protocol after acquiring the 3601 projections are the motor movements to change between the acquisition modes since mode 1 and 2 are acquired in the "front position" while mode 3 is acquired in the "mid position". This means that scanning them directly after each other would prolong the acquisition duration. To ensure that the sample mix does not move noticeably between the different acquisition modes, 10 slices have been scanned consecutively before switching the acquisition modes. 
\\[6pt]
For each acquisition mode and each 10 slice batch a dark-field and flat-field consisting of 100 averaged projections each were acquired for slice 1. Afterwards the 3,601 projections are acquired while the sample stage is rotating continuously. Subsequently, the source and detector move down by 1mm and the next 3,601 projections are acquired. This process is repeated until the 10th slice of the batch, after which also a post-batch flat-field is acquired. Then the acquisition parameters are changed for mode 2 and the above process starts again before they are changed once again for mode 3 and the process repeats one more time. After that the next 10 slice batch is scanned starting again in mode 1. A visualization of the scanning procedure can be found in Figure \ref{fig:scanning_procedure}. Depending on the available time and/or scanner errors occurring between 10 and 50 slices were acquired per scanning session. In total 5,000 slices were acquired in 111 sessions which lasted between 1h 42mins (10 slices) and 8h 34mins (50 slices).
\\[6pt]

Additionally to these slices acquired with the standard sample mix and the above mentioned acquisition parameters, the following out-of-distribution (OOD) scans were acquired: 
\begin{itemize}
    \item "pure-sample-ODD": Only one type of sample is mixed with the filler material, scanned in the same way as the standard sample mix.
    \item "foreign-objects-ODD": A new type of sample not contained in the standard sample mix is added and the resulting new mix is scanned with the standard settings. The foreign objects used are fresh figs, grapes, hazelnuts, peanuts, pistachios, and titanium prostheses screws, cf. Table \ref{tab:sample_mixes}.
    \item "Noise-OOD": the standard sample mix with a tube voltage of 90.0kV, the Thoraeus filter but an even lower tube power of 1.5W compared to the 3.0W used in mode 1. \verb|mode1| contains the 1.5W measurements, \verb|mode2| the usual noisy 3.0W and \verb|mode3| the usual "clean" 90.0W measurements.  
\end{itemize}

\textit{Remark.} During the acquisition of the dataset the detector broke down. It was exchanged by TESCAN-XRE NV and the FleX-ray scanner has been re-calibrated before resuming the dataset acquisition. Since for every 10 slice batch both dark- and flat-fields are acquired the pixel sensitivities and dark currents of the individual detector should not play a role. Table \ref{tab:slices_overview} lists which slices have been acquired on which detector and with which sample mix.

\begin{table}[ht]
\centering
\begin{tabular}{lcc}
	\hline
	\textbf{Sample} 	& \textbf{Detector 1} & \textbf{Detector 2} \\ 
	\hline
	Mix 1 (slices 1 - 1800)    & slices 1 - 1800 	& -    \\
	\hline
	Mix 2 (slices 1801 - 3720)  & slices 1801 - 2830    & slices 2831 - 3720 \\
	\hline
	Mix 3 (slices 3721 - 5000)  & -    & slices 3721 - 5000 \\
	\hline
	\hline
	Fig (OOD Pure)          & slices 5521 - 5570 	& - 	\\
	\hline
	Almond (OOD Pure)       & slices 5571 - 5620 	& - 	\\
	\hline
	Banana (OOD Pure)       & slices 5621 - 5670 	& - 	\\
	\hline
	Raisin (OOD Pure)       & slices 5671 - 5720 	& - 	\\
	\hline
	Walnut (OOD Pure)       & slices 5721 - 5770 	& - 	\\
	\hline
	Coffee beans (OOD Pure) & slices 5771 - 5820 	& - 	\\
	\hline
	Lava stone (OOD Pure)   & slices 5821 - 5870 	& - 	\\
	\hline
	\hline
	Mix 3 (OOD Noise)       & -    & slices 5871 - 5920 \\
	\hline
	Titanium prostheses screws (OOD Mix 3) & -    & slices 5971 - 6070 \\
	\hline
	Peanut (OOD Mix 3)      & -    & slices 6121 - 6170 \\
	\hline
	Pistachio (OOD Mix 3)   & -    & slices 6171 - 6220 \\
	\hline
	Hazelnut (OOD Mix 3)    & -    & slices 6221 - 6270 \\
	\hline
	Grape (OOD Mix 3)       & -    & slices 6271 - 6320 \\
	\hline
	Fresh fig (OOD Mix 3)   & -    & slices 6321 - 6370 \\
	\hline
\end{tabular}
\caption{Overview of all slices in the dataset assigned to their respective sample mix and used detector. \label{tab:slices_overview}}
\end{table}

\subsection*{Computational processing}
\label{sec:computational_processing}
The above described data acquisition process yielded a total of 540,195,000 files for the 5,000 standard slices in three modes each. The scanning of every slice produced per mode 3,601 projection data files (images of size 1 x 1912 in TIFF format) and for every 1st and 10th slice of a 10-slice-batch per mode there are either an additional pre-batch dark- and flat-field or an additional post-batch flat-field, respectively. 

\subsubsection*{Sinogram production}
\label{sec:sinogram}
To facilitate using the data collection, the 3,601 projection data files for one slice and mode were combined into one sinogram (image of size 1912 x 3601 in TIFF format) and stored in a folder together with copies of the dark- and flat-fields which belong to the respective 10-slice-batch. The script used for this (\verb|sinogram_production.py|) can be found on GitHub: \hyperlink{https://github.com/mbkiss/2DeteCTcodes}{https://github.com/mbkiss/2DeteCTcodes}.

\subsubsection*{Reconstruction production}
\label{sec:reconstruction}
The sinograms contain the raw photon counts per detector pixel that have to be corrected by off-set counts (“dark currents”) via the the dark-field (D) and pixel-dependent sensitivities via the flat-fields (F). According to the following formula the combined sinograms (S) can be corrected and converted into a beam intensity loss image (I) following the Beer-Lambert law after applying the negative logarithm to it:
\begin{equation}
    y = - \log{(I)} = - \log{ \left( \frac{S - D}{F - D} \right) } 
    \label{eq:preprocessing}
\end{equation}

The conversion of the sinograms into beam intensity loss images can in some cases yield negative or zero pixel values which were then replaced by the value $1\times10^{-6}$ to ensure strictly positive values as a pre-requisite for the subsequent negative logarithm transform. Although the filtered back-projection (FBP) is a widely used analytical technique to solve the inverse problem of CT reconstruction, noisy and beam-hardening artefact-inflicted measurements yield reconstructions with streaking artifacts. \cite{buzug2011computed}
\\[6 pt]
Therefore, the reconstructions for each slice were obtained by using an iterative reconstruction technique to solve a non-negative least squares (NNLS) problem using 100 iterations of Nesterov accelerated gradient descent \cite{nesterov1983method} with a step size of $\tau = 1/L$, where L is the Lipschitz constant of the forward operator. The forward and backward projection operators were implemented using the CUDA kernels in the ASTRA toolbox. All reference reconstructions were computed from down-scaled sinograms ($956 \times 3,601$), yielding reconstructions of $1,024 \times 1,024$ to limit the memory size of the dataset. To reduce the extent of artifacts - especially due to beam hardening in \verb|mode3| - a reconstruction plane of $\SI{233.0}{\milli \metre}^2$ was used that was resolved by $2,048^2$ pixels of physical size $\SI{113.8}{\micro \metre}^2$. After the reconstruction the images were cropped to the central square of $1,024 \times 1,024$. The computation for one reconstruction took about 63s on a NVIDIA GeForce RTX 3070 with 8GB of GDDR6 memory and an Intel Core i7-10700KF 8-core processor. The total dataset was reconstructed on a gpu-server with 4 NVIDIA GTX 1080Ti (11GB) and 2x Intel Xeon 8-core processor in 83.25 hours. Examples of reconstructed slices are shown in Figure \ref{fig:sinos_recons_segm}. The reconstructions for \verb|mode1| display a strong noise due to the limited amount of photons while image slices for \verb|mode2| appear clean, noise- and artifact-free and those for \verb|mode3| exhibit strong beam-hardening artifacts.

\subsubsection*{Segmentation production}
\label{sec:segmentation}
\begin{figure}[!ht]
	\centering
	\includegraphics[width=0.31\columnwidth]{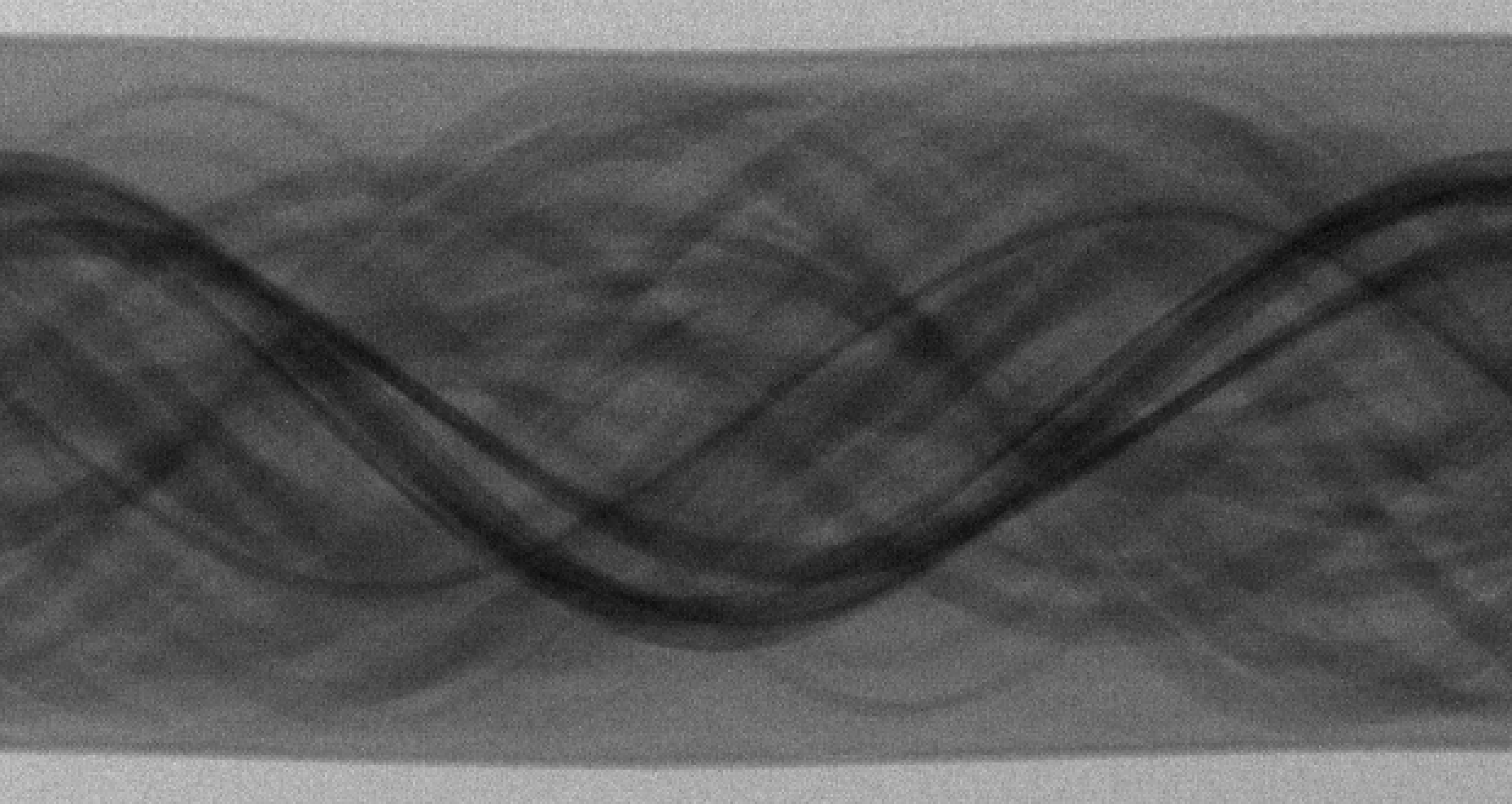}
	\includegraphics[width=0.31\columnwidth]{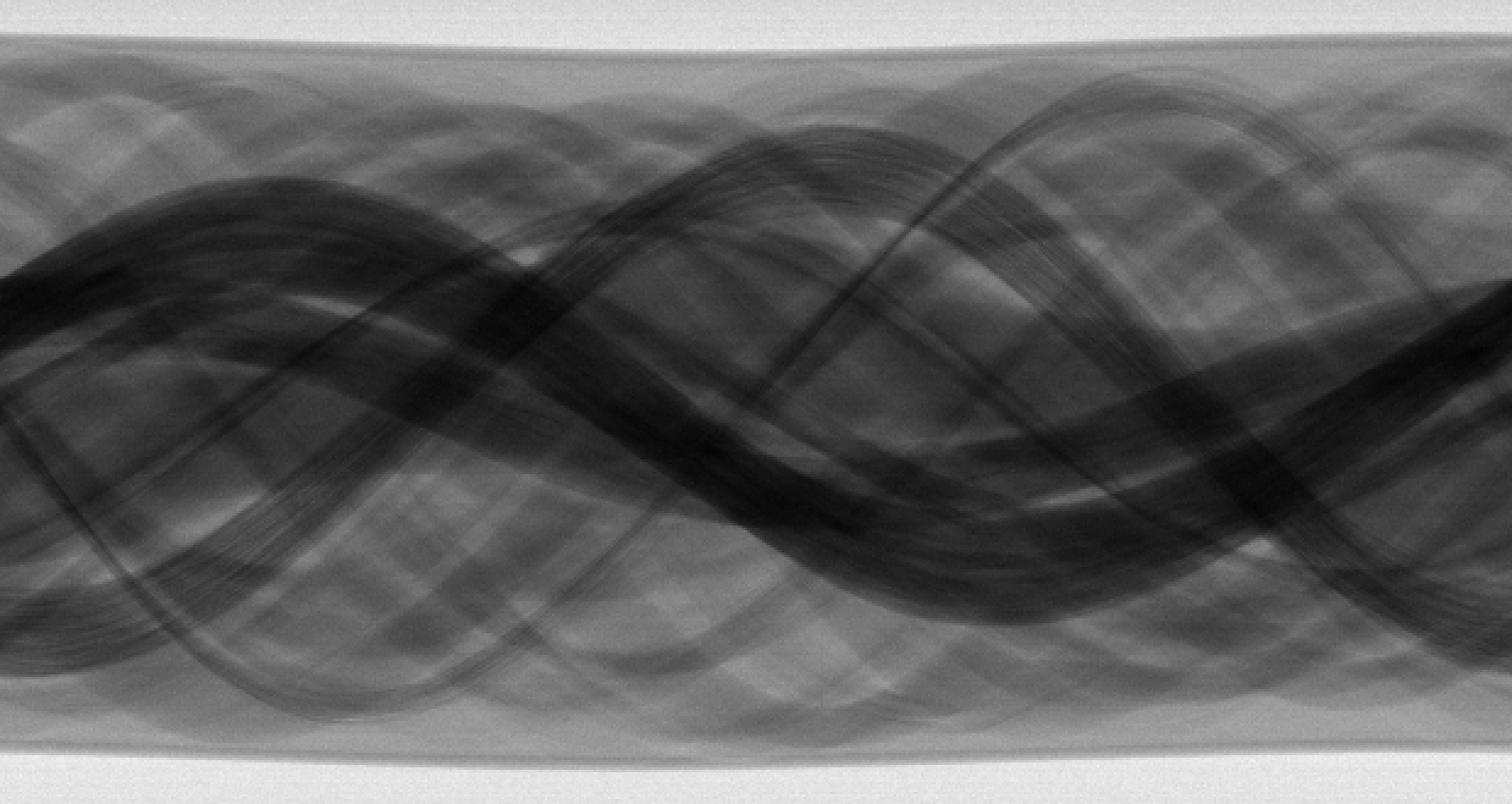}
	\includegraphics[width=0.31\columnwidth]{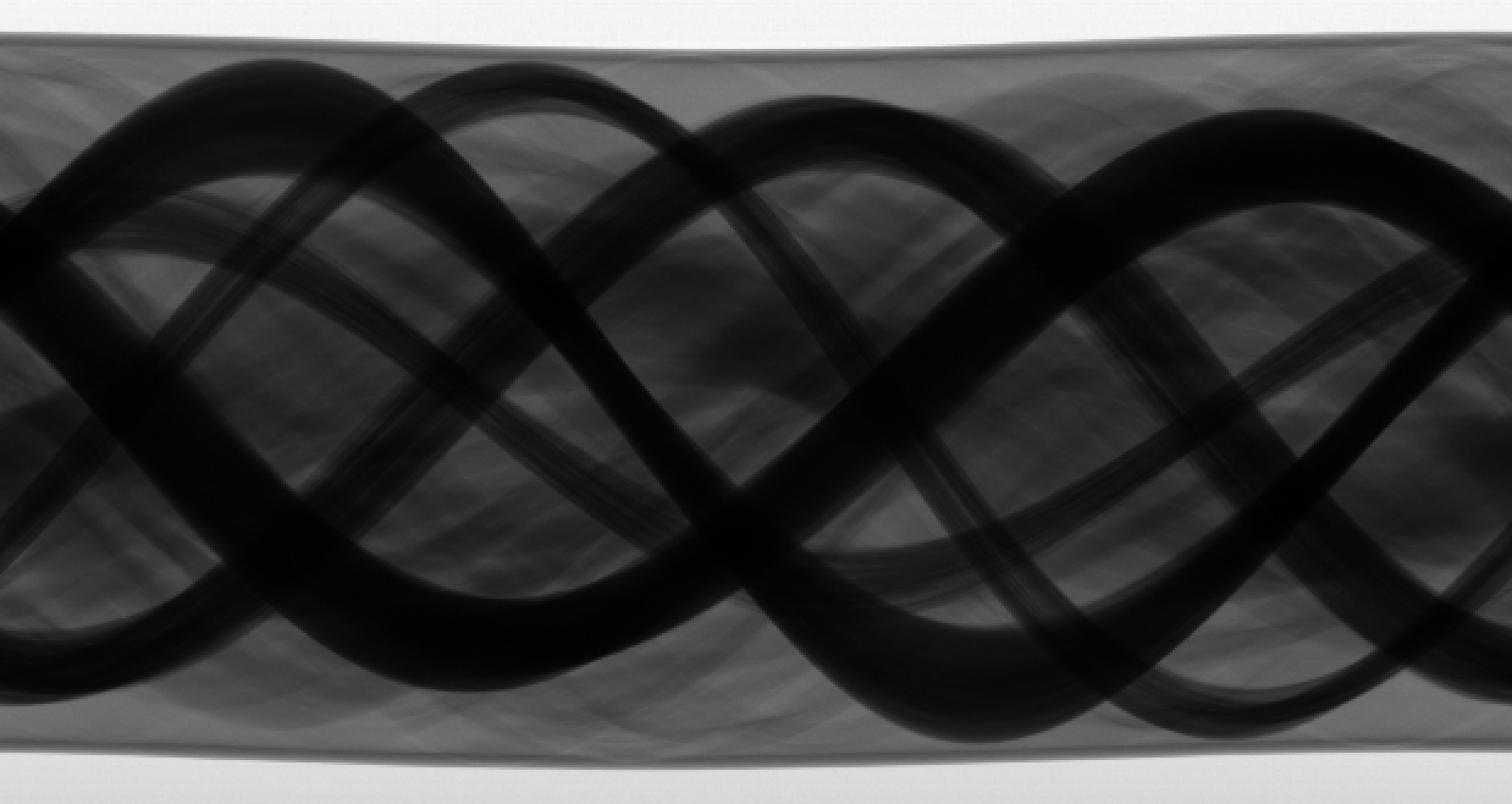}
    
	\includegraphics[width=0.23\columnwidth]{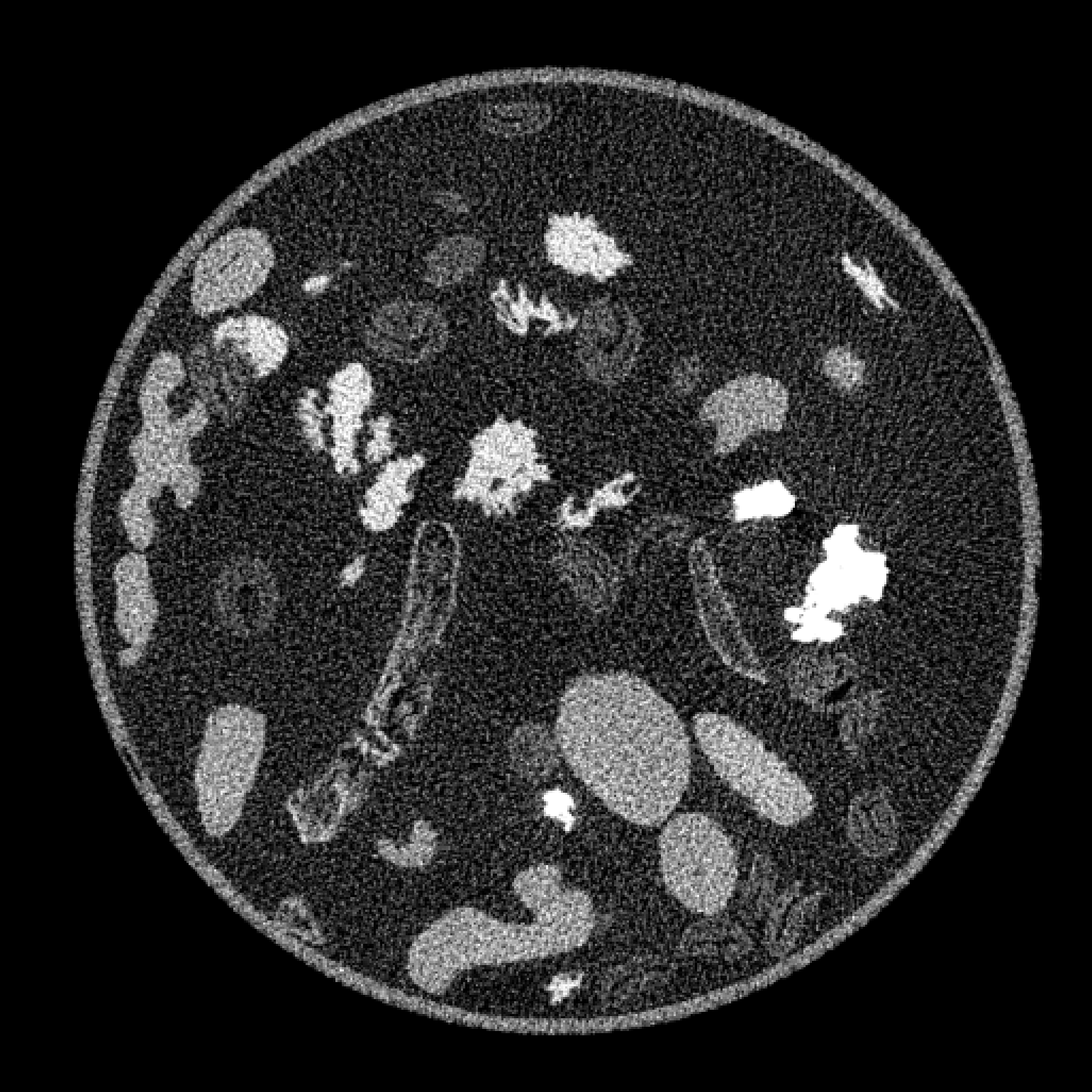}
	\includegraphics[width=0.23\columnwidth]{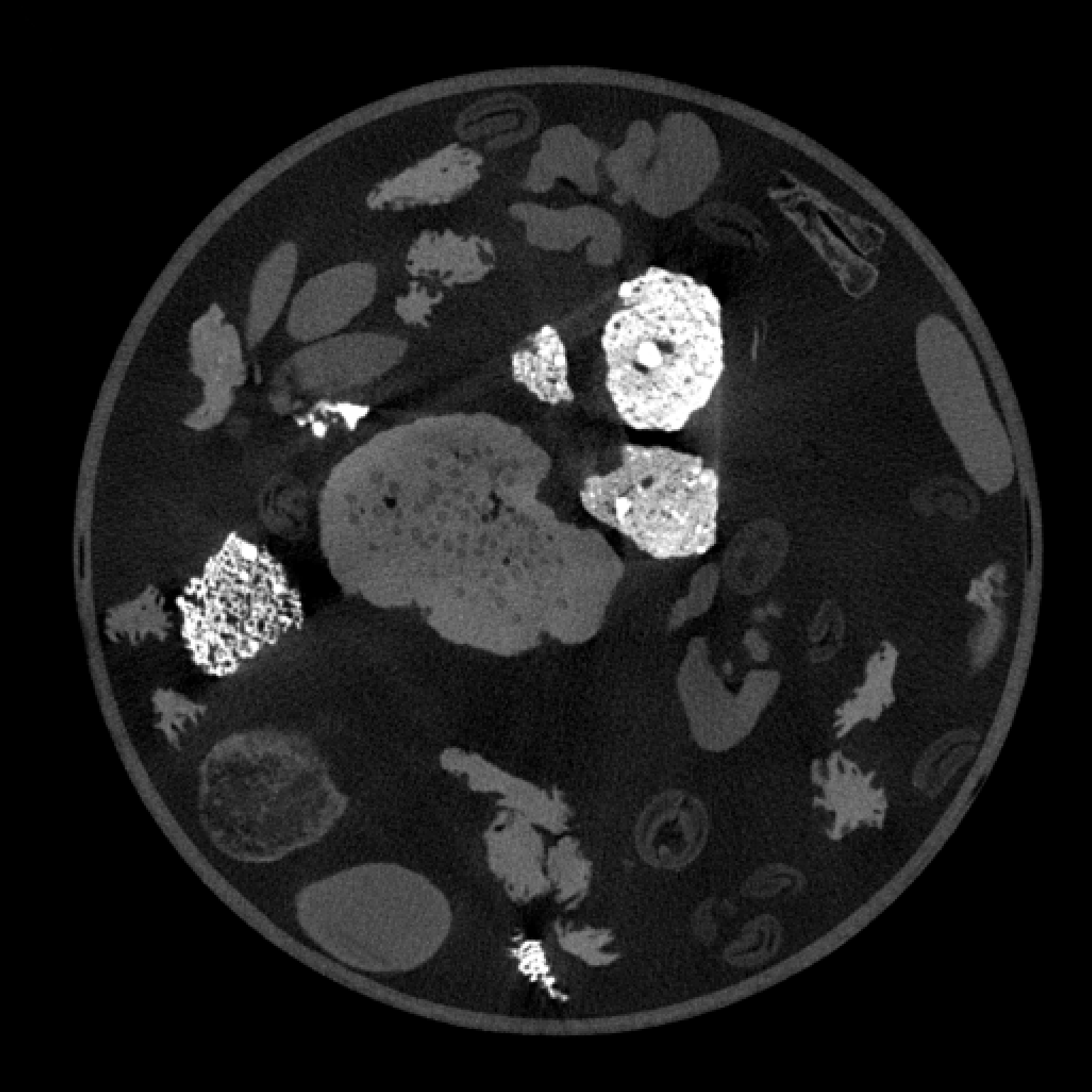}
	\includegraphics[width=0.23\columnwidth]{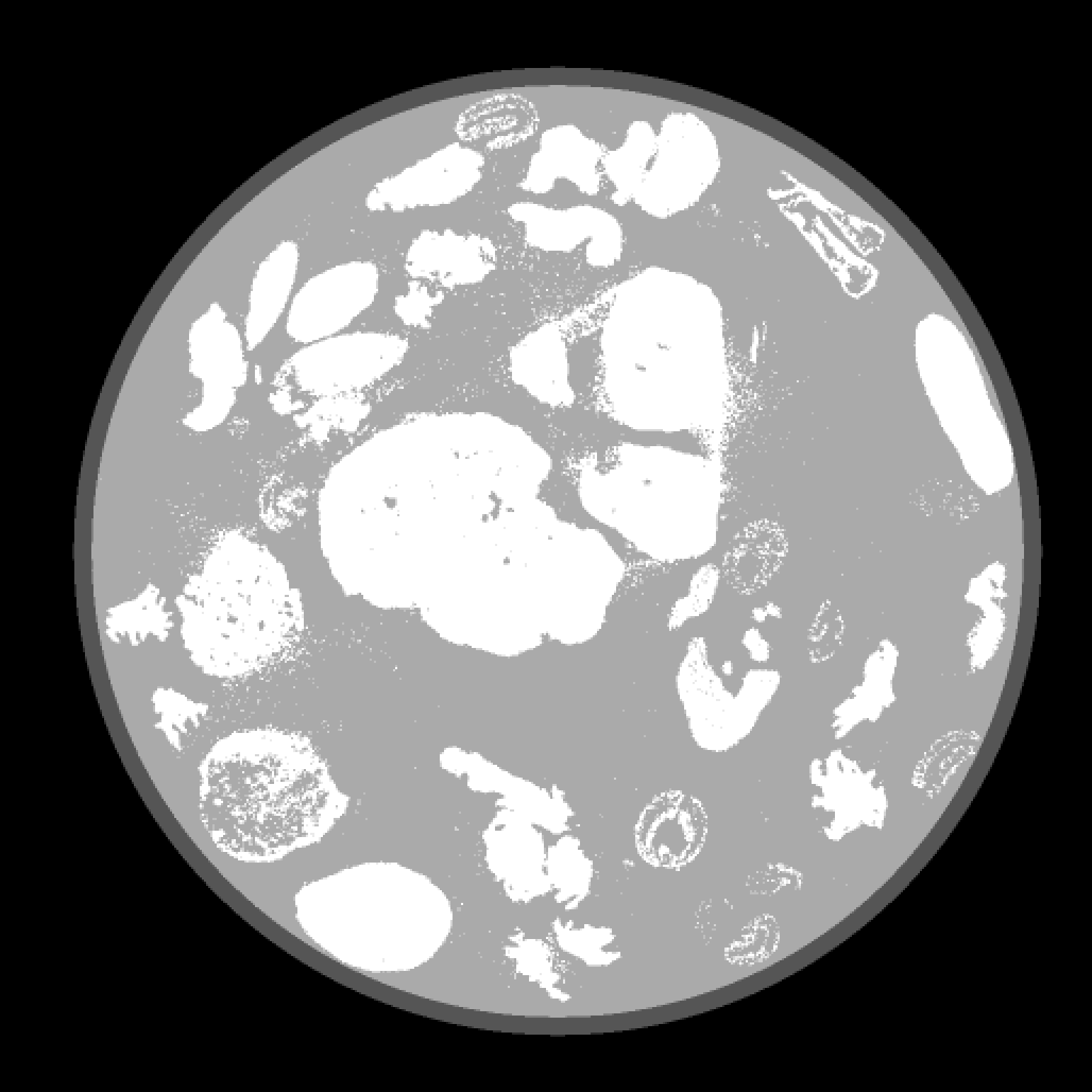}
	\includegraphics[width=0.23\columnwidth]{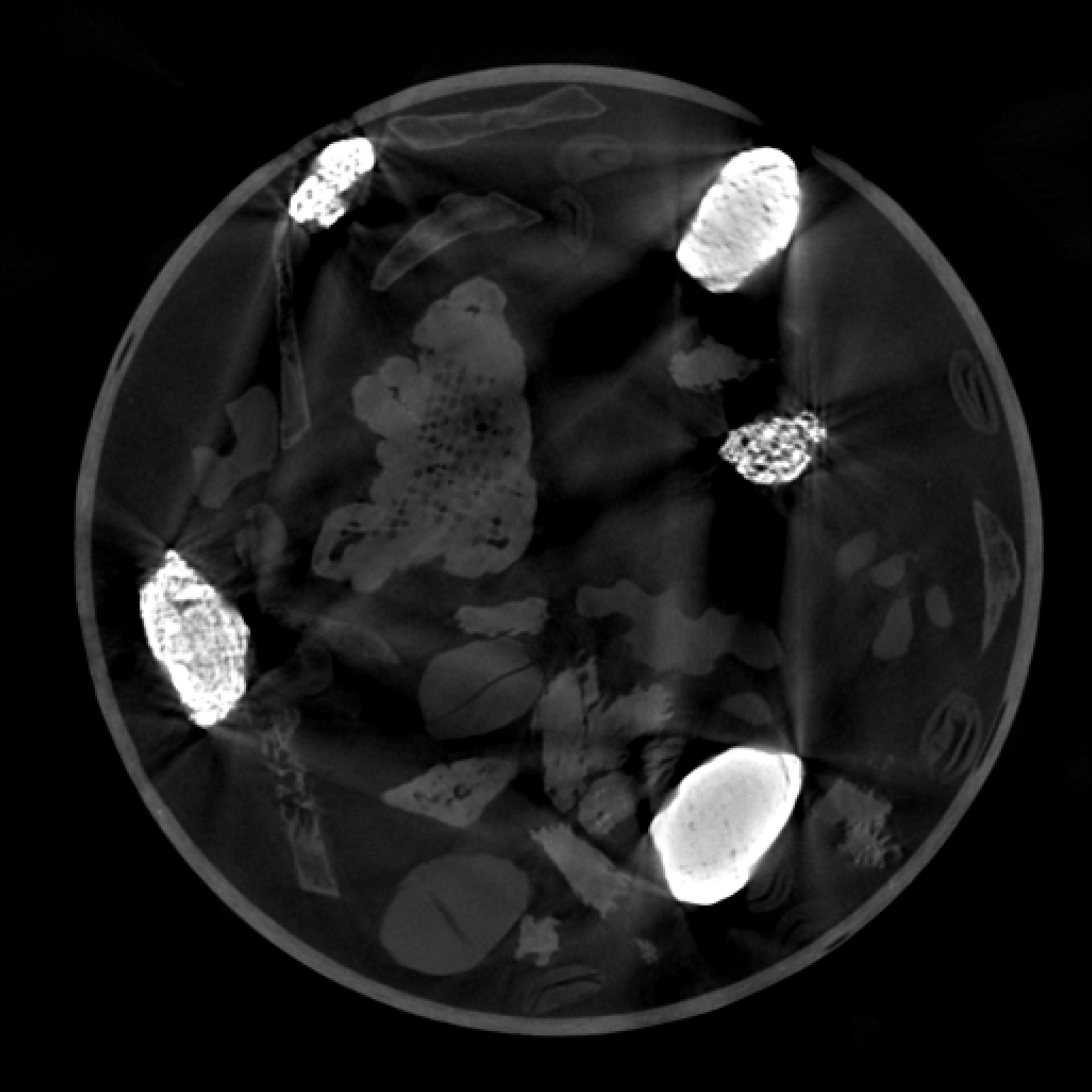}
	
	\caption{From left to right: Sinograms and reconstructions from slice 1661 (mode1 -low-dose), slice 4300 (mode2 - artifact-free), slice 560 (mode3 - artifact-inflicted) and segmentation of slice 4300 based on the mode2 reconstruction. To illustrate the variation in the sample mix we selected different slices for each mode.}
	\label{fig:sinos_recons_segm}
\end{figure}

The reference 4-class segmentation is based on the noise- and artifact-free reference reconstructions of \verb|mode2|. The segmentation distinguishes between background, tube wall, filler material and sample mix objects and is created in six steps. First, a fixed mask for the tube wall is matched to the correct location within the respective slice by maximizing the total sum of the overlap between the mask and the reconstruction via the Nelder-Mead simplex algorithm \cite{NelderMead}. Second, everything outside of the tube wall is classified as background. Third, the sample mix objects and tube wall are identified by applying a thresholding algorithm based on average thresholds [0.00110607, 0.00407358] found through a three-class multi-otsu thresholding \cite{MultiOtsu} for some sample slices. Fourth, the sample mix objects are distinguished by removing the tube wall and background from the thresholding segmentation. Fifth, the filler material is identified by removing all other classes from the whole image. Sixth, after checking that all pixels are classified once but not more than that the four-class segmentation is put together and integer values are assigned to the different classes. The values are: background - 1, tube wall - 2, filler material - 3, sample mix objects - 4. The computation for the segmentation of the total dataset took about 4.25 hours on a NVIDIA GeForce RTX 3070 with 8GB of GDDR6 memory and an Intel Core i7-10700KF 8-core processor. An example of such a four-class segmentation is shown in Figure \ref{fig:sinos_recons_segm}.

\section*{Data Records}
\label{sec:data_records}
The 2DeteCT dataset is published as open access on zenodo (\url{https://zenodo.org}) in 12 repositories. The complete data collection is organized as follows: There are a total of 5,000 slices with standard sample mix and standard acquisition parameters which are split into 10 ZIP archives containing 1,000 slices each. Additionally there are 750 OOD slices which are split into two ZIP archives. To simplify the usage of the data collection the raw projection data are split from the reference reconstructions and segmentations. Each of the raw data archives is ca. 40 GB in memory size and the corresponding reference reconstructions and segmentations amount to ca. 12 GB archives. The separate DOIs for the raw data are as follows: Slices 1 - 1,000 \cite{kiss_maximilian_b_2023_8014758}, slices 1,001 - 2,000 \cite{kiss_maximilian_b_2023_8014766}, slices 2,001 - 3,000 \cite{kiss_maximilian_b_2023_8014787}, slices 3,001 - 4,000 \cite{kiss_maximilian_b_2023_8014829}, slices 4,001 - 5,000 \cite{kiss_maximilian_b_2023_8014874}, OOD slices 5521 - 6370 \cite{kiss_maximilian_b_2023_8014907}. The reference reconstructions and segmentations can be found under these DOIs: Slices 1 - 1,000 \cite{kiss_maximilian_b_2023_8017583}, slices 1,001 - 2,000 \cite{kiss_maximilian_b_2023_8017604}, slices 2,001 - 3,000 \cite{kiss_maximilian_b_2023_8017612}, slices 3,001 - 4,000 \cite{kiss_maximilian_b_2023_8017618}, slices 4,001 - 5,000 \cite{kiss_maximilian_b_2023_8017624}, OOD slices 5521 - 6370 \cite{kiss_maximilian_b_2023_8017653}. Each slice folder \verb|slice00001| - \verb|slice05000| and \verb|slice05521| - \verb|slice06370| contains three folders for each mode: \verb|mode1|,  \verb|mode2|,  \verb|mode3|. In each of these folders there are the sinogram, the dark-field, and the two flat-fields for the raw data archives, or just the reconstructions and for \verb|mode2|  the additional reference segmentation.
\begin{itemize}
    \item \verb|sinogram.tif| is a 16-bit unsigned integer TIFF file of shape $1,912\times3,601$ containing the measured projection data of 3,601 projections acquired in a full rotation of $360^{\circ}$ combined into one sinogram (cf. Figure \ref{fig:sinos_recons_segm}); size: 13,460KB.
    \item \verb|dark.tif|  is a 16-bit unsigned integer TIFF file of shape $1,912\times1$ containing the dark-field measurement; size: 3.82KB.
    \item \verb|flat1.tif| and \verb|flat2.tif| are 16-bit unsigned integer TIFF files of shape $1,912\times1$ containing the flat-field measurements before and after every 10-slice-batch scanned; size: 3.82KB.
    \item \verb|reconstruction.tif| is a 32-bit floating point TIFF file of shape $1,024\times1,024$ containing the NNLS reconstruction computed from the pre-processed singoram according to \ref{eq:preprocessing} by 100 iterations of Nesterov accelerated gradient descent (cf. Figure \ref{fig:sinos_recons_segm}); size: 4,097KB.
    \item \verb|segmentation.tif| is 8-bit unsigned integer TIFF file of shape $1,024\times1,024$ containing the reference segmentation based on the \verb|mode2| reconstructions (cf. Figure \ref{fig:sinos_recons_segm}); size: 1,025KB.
\end{itemize}

\section*{Technical Validation}
\label{sec:technical_validation}
The FleX-ray scanner is subject to regular maintenance and calibration. Log-files of the scans have been recorded with which we can trace what happened during the scans. Furthermore, the sanity of all collected data was checked via \verb|sinogram_production.py| including that the number of files, their names and their dimensions are correct. Finally, a histogram analysis of all sinograms was performed to ensure that there is no over-saturation present.

\section*{Usage Notes}
\label{sec:usage_notes}
\subsection*{Raw projection data}
The projection data for each slice are shared as combined sinograms of 16-bit unsigned integer TIFF files containing the raw photon counts per detector pixel. TIFF files can be interpreted and manipulated by common image visualization software such as ImageJ \cite{schneider2012nih} or scientific computing languages such as MATLAB \cite{MATLAB:2010} or Python \cite{van1995python}, e.g., through the imageio package \cite{imageio}. In order to be used by tomographic reconstruction algorithms, they typically need to be pre-processed as described above and as shown in the provided scripts.

\subsection*{Reconstructions and segmentations}
In general, all reconstructions described in the previous sections can be computed from the projection data with the scripts provided. Depending on the computational resources available this could, however, require a large amount of computing time. Therefore, reference reconstructions are included in the data collection as well. They can be also used as comparison images to test novel reconstruction algorithms, or as ground truths for supervised learning algorithms. Furthermore, they can be used for CT image analysis tasks. Each reconstruction is given by a 32-bit floating point TIFF file.
\\[6pt]
The reference segmentations of the \verb|mode2| reconstructions can be computed by the users with the provided script \linebreak \verb|segmentation_production.py| as well but have been included in the uploads as well for the same reasons. Each segmentation is given by an 8-bit unsigned integer TIFF file.

\subsection*{Expansion possibilities}
Although the current scope of the 2DeteCT dataset is already offering a lot of versatile applications, the reproducible setup and experimental design as well as the availability of our highly-flexible laboratory X-ray CT scanner enable us to expand the dataset upon reasonable request. Expansion possibilities include among others: First, adding more slices with the same sample mix to increase the size of the data collection to host possible coding challenges. Second, including various new samples in the sample mix or using an entirely different sample mix for smaller expansions of the data collection. Third, adding more detailed multi-class segmentations to the dataset to train more powerful segmentation algorithms. We encourage the computational imaging community to approach us for suggestions or collaborations on such expansions.

\subsection*{Further usage}
This dataset can be used for developing both classical and ML-based algorithms for a variety of computational imaging applications, including low-dose acquisition, limited or sparse-angle scanning, beam-hardening artifact reduction, super-resolution, region-of-interest tomography or segmentation. With the provided raw projection data and the reference reconstructions in the three different modes, high-fidelity, low-dose, and beam-hardening-inflicted, it is possible to create training data pairs for supervised learning in both the sinogram and the reconstructed image domain. For example, either the A1 (low-dose) or A3 (beam hardening-inflicted) iterative reference reconstructions can be paired with the A2 high-fidelity reference reconstruction (cf. Figure \ref{fig:teaser_figure}) to train a denoiser or an artifact-reduction algorithm. For image segmentation the high-fidelity measurements or reference reconstructions of \verb|mode2| and the provided 4-class reference segmentation can be used. Alternatively, the user could use their own multi-class segmentation approach. The provided reconstruction scripts can also easily be modified by the user to represent different scan scenarios:
Limited or sparse angle tomography can be experimentally simulated by simply loading only subsets of the raw projection data; Super-resolution experiments can be conducted by either artificially binning the raw projection data into larger pixels, or by binning the reconstructed volumes into larger voxels; Region-of-interest tomography can be achieved by a combination of suitable sinogram sub-sampling and binning. In each of these cases, the provided iterative reconstruction using the full data can be set as a ground truth.

\section*{Code availability}
\label{sec:code_availability}
Python scripts for loading, pre-processing and reconstructing the projection data in the way described above are published on GitHub: \hyperlink{https://github.com/mbkiss/2DeteCTcodes}{https://github.com/mbkiss/2DeteCTcodes}. They make use of the ASTRA toolbox, which is openly available on (\hyperlink{ www.astra-toolbox.com }{ www.astra-toolbox.com }) or accessible as a conda package (\verb|conda install -c astra-toolbox astra-toolbox|). ASTRA is currently only fully supported for Windows and Linux. Installing it on Mac OS is possible but in the current state very involved and version-dependent. All reference reconstructions provided have been computed with the Python scripts. Furthermore, while the scripts allow for angular sub-sampling the projections and the reference reconstructions were computed with all projections as mentioned in the subsection "Reconstruction production" above.

\section*{Acknowledgements} 
We deeply appreciate the help of Johannes Krauß with the visualizations for this paper. We would like to thank Dr. Willem Jan Palenstijn for assisting with the computational methods. We are grateful to TESCAN-XRE NV, for their collaboration regarding the FleX-ray Laboratory. This work was supported by the Dutch Research Council (NWO, project numbers OCENW.KLEIN.285, 613.009.106, 639.073.506). The sponsors were not involved in the research and writing process.

\section*{Author contributions statement}
M.B.K. wrote the original draft of the manuscript and conceptualized the study and designed the experiments together with F.L.. M.B.K. set up all the experiments and performed the data acquisition. M.B.K. performed the data processing, inspection, as well as reorganization and wrote the reconstructions and segmentation scripts. S.B.C. wrote the scan script generator software and advised on experiments leading to this study. F.L., T.v.L., S.B.C., and K.J.B. reviewed and edited the manuscript. All authors read and approved the final manuscript. 

\bibliography{2DeteCT}

\clearpage

\section*{Appendix}

\begin{table}[ht]
\centering
\begin{tabular}{ll}
	\hline
	\textbf{Object} 	& \textbf{Density ($g/cm^3$)}\\
	\hline
	Cardboard tube          & 0.689     \\
	\hline
	Cereal coffee powder    & 0.260 	\\
	\hline
	Banana                  & 0.422 	\\
	\hline
	Coffee beans            & 0.432 	\\
	\hline
	Walnut                  & 0.494 	\\
	\hline
	Almond                  & 0.507 	\\
	\hline
	Raisin                  & 0.612 	\\
	\hline
	Fig                     & 0.629 	\\
	\hline
	Lava stone              & 1.5 - 1.9* \\
	\hline
	\multicolumn{2}{l}{*depending on its porousness of 20 – 50\% } \\
\end{tabular}
\caption{Reference densities of the final sample mix as published by the Agricultural Research Service of the U.S. Department of Agriculture (\href{https://www.aqua-calc.com}{https://www.aqua-calc.com}) \label{tab:densities_sample_mix}}
\end{table}

\begin{table}[ht]
\centering
\begin{tabular}{lcccc}
	\hline
	\textbf{Sample} 	& \textbf{\shortstack{Mix 1 \\ (slices 1 - 1,800)}} & \textbf{\shortstack{Mix 2 \\ (slices 1,801 - 3,720)}} & \textbf{\shortstack{Mix 3 \\ (slices 3,721 - 5,000)}} & \textbf{\shortstack{Mix OOD \\ (slices 5,521 - 6,370)}}\\ 
	\hline
	Cereal coffee powder    & $\SI{400}{\gram}$ 	& $\SI{400}{\gram}$    & $\SI{400}{\gram}$ & $\SI{400}{\gram}$\\
	\hline
	Banana                  & $\SI{75}{\gram}$    & $\SI{79}{\gram}$    & $\SI{80}{\gram}$ 	& $\SI{80}{\gram}$ 	\\
	\hline
	Coffee beans            & $\SI{60}{\gram}$    & $\SI{58}{\gram}$    & $\SI{54}{\gram}$ 	& $\SI{54}{\gram}$ 	\\
	\hline
	Walnut                  & $\SI{84}{\gram}$    & $\SI{73}{\gram}$    & $\SI{79}{\gram}$ 	& $\SI{79}{\gram}$ 	\\
	\hline
	Almond                  & $\SI{117}{\gram}$    & $\SI{112}{\gram}$    & $\SI{111}{\gram}$ & $\SI{111}{\gram}$ 	\\
	\hline
	Raisin                  & $\SI{111}{\gram}$    & $\SI{100}{\gram}$    & $\SI{110}{\gram}$ 	& $\SI{110}{\gram}$ 	\\
	\hline
	Fig                     & $\SI{282}{\gram}$    & $\SI{285}{\gram}$    & $\SI{290}{\gram}$ 	& $\SI{290}{\gram}$ 	\\
	\hline
	Lava stone              & $\SI{171}{\gram}$    & $\SI{154}{\gram}$    & $\SI{177}{\gram}$ 	& $\SI{177}{\gram}$ 	\\
	\hline
	\hline
	Fresh fig              & $\SI{0}{\gram}$    & $\SI{0}{\gram}$    & $\SI{0}{\gram}$ 	& 5 pieces* 	\\
	\hline
	Grape              & $\SI{0}{\gram}$    & $\SI{0}{\gram}$    & $\SI{0}{\gram}$ 	& $\SI{121}{\gram}$*	\\
	\hline
	Hazelnut              & $\SI{0}{\gram}$    & $\SI{0}{\gram}$    & $\SI{0}{\gram}$ 	& $\SI{105}{\gram}$*	\\
	\hline
	Pistachio              & $\SI{0}{\gram}$    & $\SI{0}{\gram}$    & $\SI{0}{\gram}$ 	& $\SI{87}{\gram}$*	\\
	\hline
	Peanut              & $\SI{0}{\gram}$    & $\SI{0}{\gram}$    & $\SI{0}{\gram}$ 	& $\SI{55}{\gram}$*	\\
	\hline
	Titanium prostheses screws              & $\SI{0}{\gram}$    & $\SI{0}{\gram}$    & $\SI{0}{\gram}$ 	& 4 screws*	\\
	\hline
\end{tabular}
\caption{Sample distribution for the three different sample mixes. * only one of them included in the mix during each OOD scan \label{tab:sample_mixes}}
\end{table}

\end{document}